\documentclass[]{bmcart}

\usepackage{amsthm,amsmath}
\usepackage{bm}
\usepackage[utf8]{inputenc} 
\usepackage{graphicx}
\usepackage{cancel}
\usepackage{booktabs}

\startlocaldefs

\usepackage{upgreek}
\usepackage{graphicx}
\usepackage{subfigure}
\usepackage{color}

\usepackage{color}
\usepackage[normalem]{ulem} 
\newcommand{\Replace}[2]{\bgroup\noindent\textcolor{red}{\xout{#1} #2}\egroup\ignorespacesafterend}
\newcommand{\Delete} [1]{\bgroup\noindent\textcolor{red}{\xout{#1}}\egroup\ignorespacesafterend}
\newcommand{\Insert} [1]{\bgroup\noindent\textcolor{}{#1}\egroup\ignorespacesafterend}
\newcommand{\Comment}[1]{\definecolor{Mygray}{gray}{0.50}\bgroup\color{Mygray}\noindent#1\egroup\ignorespacesafterend}
\newcommand \Michael [1]{\bgroup\noindent[\textcolor{blue}{\textbf{Michael}: #1}]\egroup\ignorespacesafterend}
\newcommand \Stefan  [1]{\bgroup\noindent[\textcolor{blue}{\textbf{Stefan}: #1}]\egroup\ignorespacesafterend}
\newcommand{\Rho}{\mathrm{P}} 

\DeclareMathAlphabet{\Ibb}{U}{msb}{m}{n}

 \newcommand{\BM}{{\boldsymbol{\mathnormal M}}}

 \newcommand{\Bsigma} {\ensuremath{\boldsymbol\sigma}}

 \newcommand{\Bve    }{\ensuremath{\boldsymbol\varepsilon}}

\newcommand{\Bb}{{\boldsymbol{\mathnormal b}}}

\newcommand{\Bn}{{\boldsymbol{\mathnormal n}}}

\newcommand{\Br}{{\boldsymbol{\mathnormal r}}}
\newcommand{\Bs}{{\pmb{\mathnormal s}}}
\newcommand{\Bt}{{\boldsymbol{\mathnormal t}}}

\newcommand{\BSigma  }{\ensuremath{\boldsymbol\Sigma}}

\newcommand{\figref}[1]{Fig.~\ref{#1}}

\newcommand \MZ [1] {\bgroup\noindent[\textcolor{blue}{\textbf{MZ}: #1}]\egroup\ignorespacesafterend}

\newcommand \Mdel [1] {\bgroup\noindent[\textcolor{red}{\textbf{Mdel}: #1}]\egroup\ignorespacesafterend}
\endlocaldefs

\newcommand \Madd [1] {\bgroup\noindent[\textcolor{blue}{\textbf{Madd}: #1}]\egroup\ignorespacesafterend}
\endlocaldefs

\begin{document}

\begin{frontmatter}

\begin{fmbox}
\dochead{Research}


\title{Relating plasticity to dislocation properties by data analysis: scaling vs. machine learning approaches}


\author[
   addressref={aff1},                   
]{\inits{SH}\fnm{Stefan} \snm{Hiemer}}
\author[
   addressref={aff2},                   
]{\inits{HF}\fnm{Haidong} \snm{Fan}}
\author[
   addressref={aff1},
   corref={aff1},                       
   email={michael.Zaiser@fau.de}
]{\inits{MZ}\fnm{Michael} \snm{Zaiser}}


\address[id=aff1]{
  \orgname{Department of Materials Simulation, Friedrich-Alexander Universität Erlangen-Nürnberg}, 
  \street{Dr.-Mack-str. 77},                     %
  \postcode{90762}                                
  \city{Fürth},                              
  \cny{Germany}                                    
}
\address[id=aff2]{
  \orgname{Department of Mechanics, Sichuan University}
  \city{Chengdu},                              
  \cny{P.R. China}                                    
}


\begin{artnotes}
\end{artnotes}

\end{fmbox}


\begin{abstractbox}

\begin{abstract} 
Plasticity modelling has long been based on phenomenological models based on ad-hoc assuption of constitutive relations, which are then fitted to limited data. Other work is based on the consideration of physical mechanisms which seek to establish a physical foundation of the observed plastic deformation behavior through identification of isolated defect processes ('mechanisms') which are observed
either experimentally or in simulations and then serve to formulate so-called physically based models. Neither of these approaches is adequate to capture the complexity of plastic deformation which belongs into the realm of emergent collective phenomena, and to understand the complex interplay of multiple deformation pathways which is at the core of modern high performance structural materials. Data based approaches offer alternative pathways towards plasticity modelling whose strengths and limitations we explore here for a simple example, namely the interplay between rate and dislocation density dependent strengthening mechanisms in fcc metals. 
\end{abstract}


\begin{keyword}
\kwd{Dislocation plasticity}
\kwd{Scaling invariance}
\kwd{Machine learning}
\end{keyword}


\end{abstractbox}
%

\end{frontmatter}



\section[Introduction]{Introduction}

Plasticity modelling has long proceeded along two independent lines. On the one hand, engineers are seeking tools to predict the behavior of engineering components during processing and under in-service loads. In this case, the material is assumed as given and the task is to predict, based on available experimental data, as accurately as possible the behavior of this material within complex-shaped parts with equally complex boundary loadings. This has led to phenomenological models which seek, often with an abundance of parameters, to reproduce a set of experimental data as accurately as possible. The task of these models is to reproduce the behavior of a given material under a typically rather limited range of deformation conditions, and in meeting this task they often achieve an impressive degree of accuracy. Their predictive power beyond the material and range of deformation conditions for which they have been parametrized, on the other hand, is extremely limited. Therefore, materials scientists tasked with the development of new and improvement of existing materials tend to use a different approach. By analysing deformation on the level of the defect microstructure, they seek to identify the physical mechanisms that control macroscopic features of the plastic deformation process such as the flow stress. A mathematical description of the corresponding microstructure-property relationships, so it is hoped, may provide generic insights that can be used as a basis for predictive modelling. But this approach, which is supported by microstructure characterization tools of increasing sophistication, is itself beset by pitfalls.

On a most elementary level, the flow stress of a dislocated crystal can be related to the density and arrangement of crystal lattice dislocations. This is well established since the seminal paper of Taylor \cite{taylor1934mechanism} who analyzed crystal plasticity in terms of the motion of dislocations and established a fundamental relationship between the flow stress $\tau_{\rm f}$ -- here understood as critical resolved shear stress on the active slip system(s) -- and the density of dislocations $\rho$. , $\tau_{\rm f} = \alpha \mu b \sqrt{\rho}$ where $\mu$ is the shear modulus, $b$ the length of the Burgers vector of the active slip systems, and $\alpha$ a numerical factor which was put by Taylor into the range of $\alpha \approx 0.2...0.3$ where it has remained ever since. However, while the structure of the Taylor relationship has never been in question, the precise nature of the dislocation arrangements that give rise to this dependency has been a subject of controversy for decades. Taylor considered a checkerboard pattern of positive and negative dislocations ('Taylor lattice') - an arrangement which, while analytically tractable, has the disadvantage that it has never been observed in experiment. Based on surface observations (blocking of slip lines), Mott and Seeger \cite{mott1953bakerian,seeger1957work} proposed that the Taylor stress is produced by large pile ups of dislocations at indestructible Lomer-Cottrell Barriers, an idea which, while consistent with surface observations, cannot be reconciled with TEM where such pile ups are very hard to find. Bayley and Hirsch \cite{bailey1960dislocation} instead suggested that Taylor's law can be explained by the stress needed to cut forest dislocations, whose spacing scales like the square root of dislocation density. Finally, Hirsch and Warrington \cite{hirsch1961flow} pointed out that Taylor-type behavior can also be explained by the dragging of jogs whose spacing in turn reflects the forest spacing. 

The controversies surrounding the mechanism of hardening illustrate an inherent weakness in the quest for 'mechanism-based' interpretations of complex collective phenomena. Virtually all of the above mentioned dislocation configurations and mechanisms (with possible exception of the original Taylor lattice) can be observed in TEM imaging, but possible selection bias makes it difficult to quantify their relevance based on published data. Another example of the problematics of mechanistic thinking concerns the nature and role of dislocation sources. Every dislocations textbook contains images of Frank-Read or spiral sources, and numerous attempts to explain the flow stress of materials from the macro to the micro scale are based on the concept of a 'weakest source' Yet, while dislocation sources can be observed in TEM, they are surprisingly rare and the actual process of dislocation multiplication does {\em not} proceed by seqential emission of discrete loops from dislocation sources but in a much more diffuse and mechanistically less tangible manner which, despite recent works such as the excellent simulation study of Weygand and co-workers \cite{stricker2018dislocation}, is still not fully understood. Thus, even simple and elegant mechanisms on the single-dislocation level do not necessarily help to obtain an adequate understanding of the inherently collective and complex dynamics of dislocation networks.

Here we illustrate two alternative approaches towards quantifying the relationship between plastic deformation behavior and properties of the dislocation microstructure. High-throughput simulations and experimentation offer the perspective to establish a sound data base which allows data analytic methods to be used for identification and classification of recurrent features and structures. Mathematical analysis allows to formulate symmetries and invariance principles that reduce data complexity and assist in analysis. In the present study, we illustrate these approaches on a simple example, namely the superposition of rate and dislocation density effects in controlling the flow stress of fcc single crystals.

\section{Data base }

High throughput discrete dislocation dynamics (DDD) simulations covering 9 orders of magnitude in dislocation density $\rho$ (10$^{7}$m$^{-2} < \rho < 10^{16}$m$^{-2}$   and 7 orders of magnitude in strain rate ($10^{-1}$s$^{-1} \le
\dot{\varepsilon} \le < 10^{6}$s$^{-1}$) were conducted by Fan and co-workers \cite{fan2021strain} with the aim of establishing the joint influence of dislocation density and strain rate on the flow stress of fcc metals. These simulations were complemented by MD simulations of highly dislocated crystals which further extend the range of dislocation densities and strain rates to densities of $2.2  \times 10^{16}$ m$^{-2}$ and strain rates of $2.5\times 10^8$ s$^{-1}$. For each set of parameters multiple simulations with different, but statistically equivalent, initial dislocation configurations were conducted, amounting to a total of about 200 simulations. In all these simulations the flow stress, defined as the stress at a fixed plastic strain $\varepsilon^{\rm p}_{\rm y}$, was recorded alongside the imposed strain rate and dislocation density at the same strain.  A default value $\varepsilon^{\rm p}_{\rm y} = 0.5\%$ was used for the offset plastic strain, though lower offsets were considered at the lowest strain rates. In addition to flow stress values, other characteristics such as the probability distribution $p(v)$ of dislocation segment velocities and the plastic strain pattern at the global strain $\varepsilon^{\rm p}_{\rm y}$ were determined for all simulations. 

The simulations were complemented by an extensive literature search to retrieve records where simultaneous measurements of strain, strain rate, flow stress and dislocation density were reported for monocrystalline speciens. This search yielded about 120 datasets, mostly from the literature of the 1960s to 1980s. Few examples from the recent literature could be found, since unfortunately the number of publications which report quantitative measurements of dislocation densities alongside mechanical data has over the past decades decreased in inverse proportion with the increasing number of plasticity models that use dislocation densities as internal variables. A downloadable compilation of all data can be found in the supplementary material of Ref \cite{fan2021strain}.  

In the following we investigate the performance of different prediction strategies in relating the flow stress to parameters such as strain, dislocation density and strain rate. As a measure of prediction performance we use the prediction score
\begin{equation}
	R^2 = 1 - \frac{\sum_i (x_i - x_i^{\rm pred})^2}{\sum_i (x_i - \langle x_i \rangle^2}
\end{equation}
which we apply to the logarithm of the flow stress, which we try to predict based on the remaining variables (note that in view of the range of variation of all variables, which encompasses many orders of magnitude in disloation density, strain rate and flow stress, a logarithmic measure is required).

\section{Predicting flow stresses: Superposition of forest hardening and rate effects}

We first consider a simple question: How do dislocation interactions and rate-dependent flow stress contributions, which can ultimately be traced back to the stress needed to move dislocations with a given imposed velocity, superimpose in controlling the flow stress of fcc metals? In other words, what is the function that relates the flow stress $\sigma_{\rm f}$ to dislocation density $\rho$ and strain rate $\dot{\epsilon}$? In the literature, there exists an abundance of phenomenological relationships introduced by different authors in a more or less ad-hoc manner. In particular we mention the form popularized by Mecking and Kocks \cite{mecking1981kinetics},
\begin{equation}
\sigma_{\rm f} = \left(\frac{\dot{\epsilon}}{\dot{\epsilon}_0}\right)^{1/m} \alpha_0 \mu b \sqrt{\rho}
\label{eq:kocksmecking}
\end{equation}
where $\dot{\epsilon}_0$ is an arbitrary reference strain rate, $\alpha_0$ an accordingly determined nondimensional factor, $\mu$ the shear modulus, and $b$ the Burgers vector length. The exponent $m$ is called the strain rate sensitivity. We shall probe the usefulness of this and similar equations in reproducing data based on a combination of theoretical and data based analysis. (For a discussion of other aspects, such as thermodynamic consistency, see \cite{wu2022thermodynamic}). 

\subsection{Scaling analysis}

In the following we show how to exploit generic scaling invariance properties of dislocation systems in order to establish constraints on the possible form of constitutive laws that connect statistically averaged properties of dislocation systems such as stress, strain rate, and dislocation density. The main arguments were for the first time formulated by Zaiser and Sandfeld \cite{zaiser2014scaling}
and are here briefly repeated. 

While it is uncommon to envisage discrete dislocation dynamics simulations in terms of equations of motion for the dislocation lines, this is of course possible in terms of a balance of local shear stresses acting on the dislocation lines. We consider a system of $N$ dislocations  $i\in\{1\ldots N\}$. The dislocation with Burgers vectors $\Bb^{\,i}$ moves by glide on the slip plane ${\cal P}^{i}$ with slip plane normal $\Bn^{i}$. The unit slip vector is $\Bs^i = \Bb^{\,i}/b$ where $b$ is the Burgers vector length. The dislocations form closed loops ${\cal C}^i$ contained within a single slip plane. These loops are  parameterized by $\Br(s^i)$ with local tangent vector $\Br(s^i) = {\rm d}\Br/{\rm d} s^i$. Junctions are described in terms of local alignment of segments of different loops. We consider bulk behaviour, i.e., we  assume that the dislocation loops are contained within a quasi-infinite crystal where the boundaries are remote such that image stresses can be neglected, or that the system is replicated periodically. 
 
Dislocation motion is assumed to occur by glide and to be controlled by phonon drag with drag coefficient $B$. Thus, the local velocity is given by
\begin{equation}
	\label{eq:motion}
	\frac{\partial \Br(s^i)}{\partial t} = [\Bn^i \times \Bt(s^i)] v(s^i)\quad,\quad
	v(s^i) = \frac{b}{B} \tau^i(s^i), \quad,\quad \tau^i(s^i) = \BM^i:\Bsigma(\Br(s^i))
\end{equation}
where $\BM^i = [Bn^i \otimes \Bs^i]^{\rm sym}$ is the slip system projection tensor. The stress is composed of an 'external' stress imposed by remote boundary tractions and considered constant over the volume of interest, and a dislocation-related internal stress which can be computed in terms of line integrals over the dislocation lines: 
\begin{eqnarray}
	\Bsigma &=& \Bsigma^{\rm ext} + \Bsigma^{\rm int}\nonumber\\
	\sigma_{kl}^{\rm int}(\Br) &=& -\frac{\mu}{8\pi} \sum_k \int_{{\cal C}^k}\left\{  \frac{2}{1-\nu}\left( \frac{\partial^3 R}{\partial r_n \partial r_k \partial r_l} - \delta_{kl} \frac{\partial}{\partial r_n}\nabla^2R \right) b_o\epsilon_{nom}t_m \right. \nonumber\\
	&+& \left.\left(\frac{\partial}{\partial r_n} \nabla^2 R\right) b_o  \left[\epsilon_{nok}\, t_l    + \epsilon_{nol}\, t_k \right] \right\}{\rm d}s^k, 
	\label{eq:3Dstress}
\end{eqnarray}
where $R = |\Br - \Br(s^k)|$. 
It is now easy to see that the above equations are invariant upon the re-scaling 
\begin{equation}
	\label{eq:trans}
	\Br  \to \lambda \Br\quad,\quad
	\Bsigma \to \lambda^{-1} \Bsigma \quad,\quad
	t \to \lambda^{n+1} t \quad.
\end{equation}
which implies the auxiliary transformations
\begin{equation}
	\label{eq:trans2}
	v \to \lambda^{-1} v \quad,\quad
	\rho \to \lambda^{-2}\rho \quad,\quad
	\dot{\Bve}^{\rm p} \to \lambda^{-3} \dot{\Bve}^{\rm p}  \quad.
\end{equation}
where the transformation rule for dislocation density follows directly from its as line length per volume. The transformation rule for the plastic strain rate follows from its definition in terms of slip rates on the different slip systems, $\dot{\Bve}^{\rm p} = \sum_{\beta}\BM^{i}\dot{\gamma}^{i}$, where the slip rates $\dot{\gamma}^{i} = b \dot{A}^{i}/V$ are products of Burgers vector length and rate of change in slipped area per dislocation loop, divided by the system volume. 


\subsubsection{Crystal plasticity constitutive equations}

We now outline some consequences of the above formulated invariance principle. It is clear that invariance under the transformation (\ref{eq:trans} )does not depend on any mechanisms or specific processes, nor does it depend on the scale on which the dislocation system is considered. Scaling invariance must not only hold on the macroscopic scale, but must also apply to any emergent statistical signatures of the evolving dislocation system. This is in well known for the characteristic wavelength of dislocation patters, which obeys the so-called law of similitude \cite{rudolph2005scaling,sauzay2011scaling} as well as for the mesh length distribution in fractal dislocation networks \cite{zaiser1999flow,hahner1999} and the distribution of dislocation velocities \cite{fan2021strain}.  In particular, any constitutive equations that derive from the micro-dynamics of interacting dislocations via an averaging procedure are bound to possess the same invariance properties. This provides us with a useful rule-of-thumb for assessing the validity of phenomenological dislocation-based constitutive equations proposed in the literature. For instance, simple power counting demonstrates that Eq. (\ref{eq:kocksmecking}) is invariant under (\ref{eq:trans}) only in the rate independent limit $m \to \infty$, and can therefore not meaningfully describe rate effects in dislocation plasticity.

Conversely, a useful strategy for formulating constitutive equations is to cast these in the form of relationships between {\em invariant parameters} that by construction show invariance under the transformation (\ref{eq:trans}). We demonstrate this strategy for the superposition of rate and dislocation density effects in dislocation plasticity. Thus, we define invariant slip rate and dislocation density variables on the different slip systems $\beta$ via
\begin{equation}
	\Rho^{\beta} = \left(\frac{\mu b^3}{B}\right)^{2/3}\frac{\rho^{\beta}}{(\dot{\gamma}^{\beta})^{2/3}}
	\quad,\quad
	\dot{\Gamma^{\beta}} = \frac{B}{\mu b^3}\frac{\dot{\gamma}}{(\rho^{\beta})^{3/2}} = (\Rho^{\beta})^{-3/2}
\end{equation}
Similarly, we define invariant stress and strain measures via 
\begin{equation}
	\BSigma_{\rho} = \frac{\Bsigma}{\mu b \sqrt{\rho}}
	\quad,\quad
	\BSigma_{\dot{\gamma}} = \frac{\Bsigma}{\mu^{1/3}(B\dot{\gamma})^{2/3}}
	\quad,\quad
	\Gamma^{\beta} = \frac{\gamma^{\beta}}{b\rho^{1/2}}. 
\end{equation}
from which invariant shear stresses derive via 
$$
	T_{\rho}^{\beta} = \BM^{\beta}:\BSigma_{\rho}
	\quad,\quad
	T_{\dot{\gamma}} = \BM^{\beta}:\BSigma_{\dot{\gamma}}. 
$$
Note that we have non-dimensionalized all quantities using the material constants which govern dislocation motion and interactions, in order to allow for a material independent formulation of constitutive behavior associated with collective dynamics of dislocations. 

Next we study asymptotic cases. We note that the slip rates scale on the active slip systems $\beta$ are given by $\dot{\gamma}^{\beta} = \rho^{\beta} b v^{\beta}$ where $\rho^{\beta}$ is the dislocation density on a given slip system and $v^{\beta} = \dot{\gamma}^{\beta}/(\rho^{\beta} b)$ the average velocity of these dislocations.  

First we envisage the quasistationary limit $v^{\beta} \to 0$ of near-zero strain rates or of very high dislocation densities. In this limit the stresses on all dislocation lines asymptotically vanish, $\Bsigma^{\rm ext} + \Bsigma^{\rm int} \to 0$, from which scaling invariance dictates that $\tau^{\beta} \propto \sqrt{\rho}$ for all active slip systems $\beta$. In terms of the invariant variable $\Rho^{\beta}$ this behavior is expressed as
\begin{equation}
	T^{\beta}_{\dot{\gamma}} = \alpha^{\beta} (\Rho^{\beta})^{1/2}  \quad,\quad
	T^{\beta}_{\rho} = \alpha^{\beta} \quad,\quad
	\Rho^{\beta} \to \infty, \dot{\Gamma}^{\beta} \to 0. 
\end{equation}
where the parameters $\alpha^{\beta}$ may depend on the distribution of dislocations over the different slip systems as expressed by the ratios $f^{\beta} = \rho^{\beta}/\rho$. 

In the opposite limit $v^{\beta} \to  \infty$ of low dislocation densities or high strain rates, the resolved shear stresses acting on the dislocations must become very high. This is only possible when the externally applied stresses are high and in the asymptotic limit, the internal stresses are asymptotically irrelevant. Thus, in the asymptotic limit $v^{\beta} = \dot{\gamma}^{\beta}/(\rho^{\beta} b) = \tau^{\beta} B/b$. From this relation we obtain 
\begin{equation}
	T^{\beta}_{\dot{\gamma}} = (\Rho^{\beta})^{-1} \quad,\quad
	T^{\beta}_{\rho} = (\Rho^{\beta})^{-3/2} \quad,\quad
	\Rho^{\beta} \to 0, \dot{\Gamma}^{\beta} \to \infty. 
\end{equation}
An equivalent formulation is of course possible by substituting $\Rho^{\beta} = (\dot{\Gamma}^{\beta})^{-3/2}$. A generic constitutive law must interpolate between the asymptotic limits given above. The simplest way to do so is to simply add up the asymptotic expressions, as proposed by Fan et.al. \cite{fan2021strain}:  
\begin{equation}
T^{\beta}_{\dot{\gamma}} = (\Rho^{\beta})^{-1} + \alpha^{\beta} =   (\Rho^{\beta})^{1/2}\quad,\quad
T^{\beta}_{\rho} = \alpha^{\beta} + (\Rho^{\beta})^{-3/2} = \alpha^{\beta} +  \dot{\Gamma}^{\beta}.
\end{equation}
\begin{figure}[tb]
	\centering
	\hbox{}
	\includegraphics[width=0.48\textwidth]{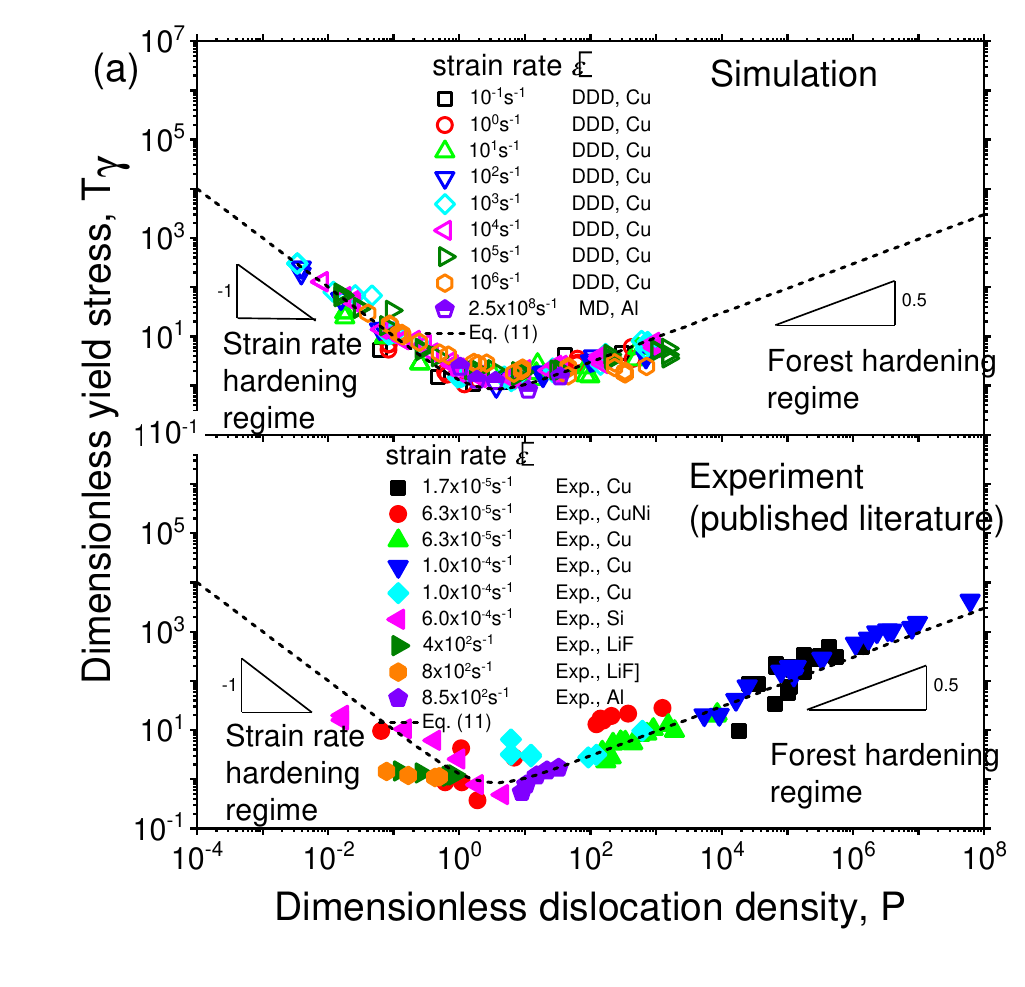}\hfill
	\includegraphics[width=0.48\textwidth]{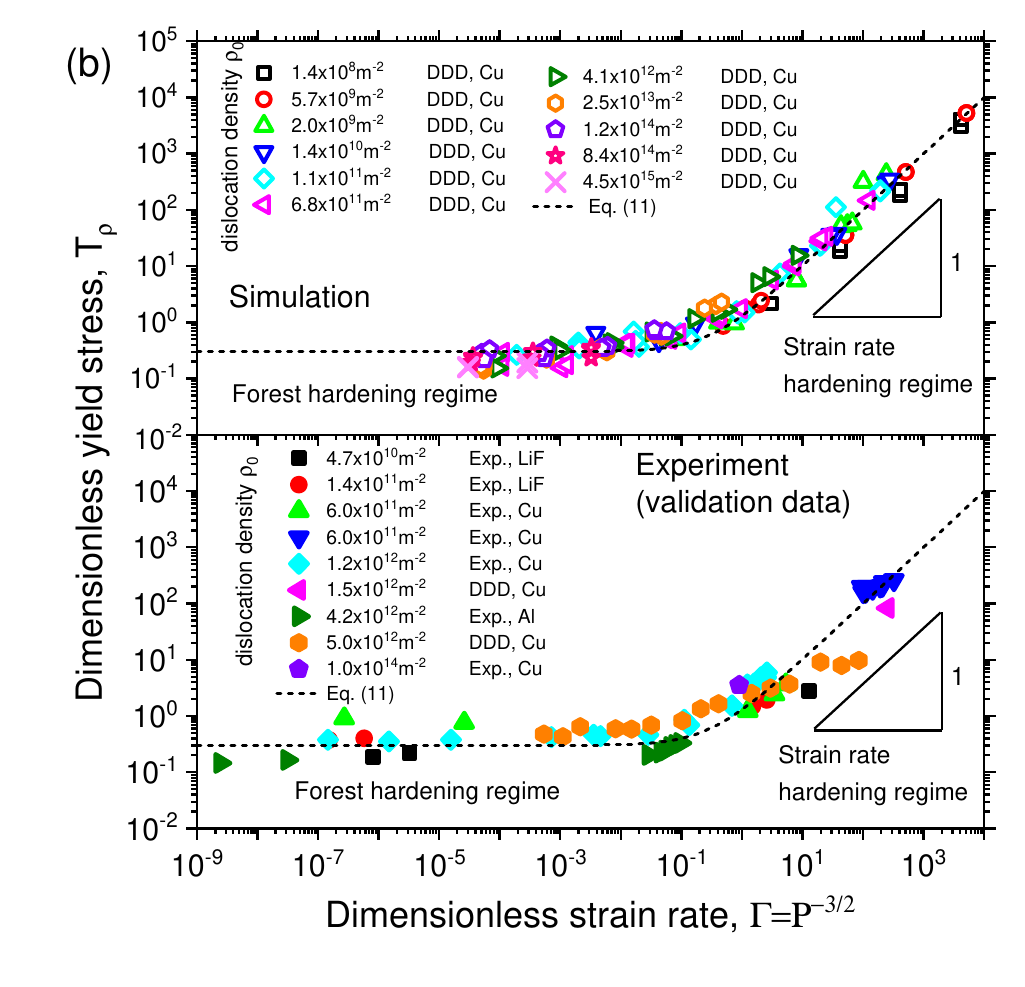}
	\caption{\label{fig:comparison}
		Comparison between scaling function and data describing the dependency of flow stress on strain rate and dislocation density, figure reproduced from Ref. \cite{fan2021strain}.}
\end{figure}
\begin{figure}[b]
	\centering
	\hbox{}
	\includegraphics[width=0.48\textwidth]{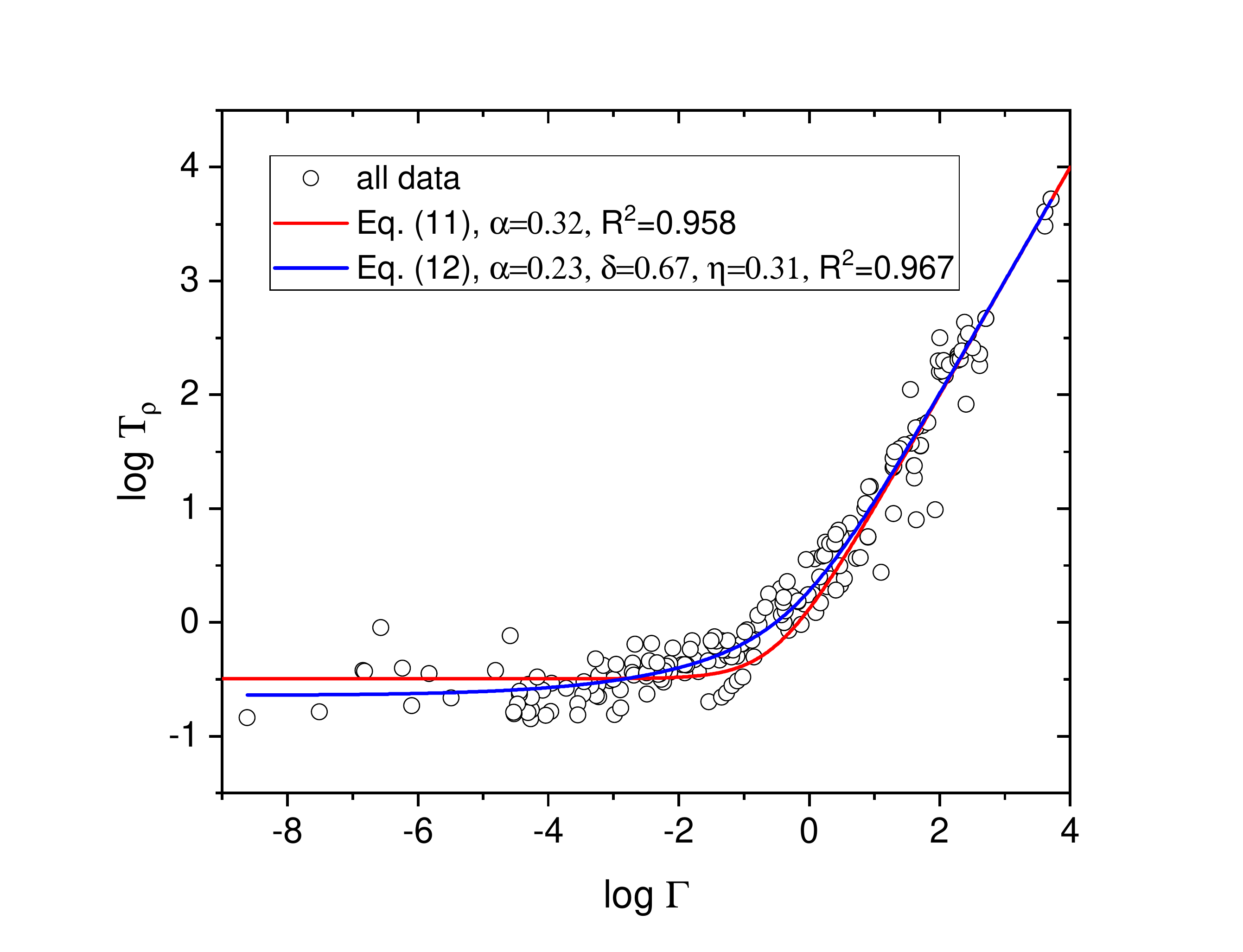}
	\caption{\label{fig:fitting}
		Fitting the data of \figref{fig:comparison} with a modified law corresponding to a different asymptotic strain rate exponent; blue curve: best fit according to Eq. (11), blue curve: best fit according to Eq. (13), fit parameters see legend.}
\end{figure}
Eq. (11) is compared in \figref{fig:comparison} with the data of Fan et. al. \cite{fan2021strain}, who in their simulations consider uniaxial tensile tests of Cu and Al deformed in [100] orientation, and with experimental data referring to uni-axial tests compiled from the literature \cite{fan2021strain} which have been corrected for orientation factors. It can be seen that the simple relationship (11) provides an acceptable description of the data. We measure prediction performance in terms of the coefficient of determination
\begin{equation}
	R^2_{\tau} = 1  \frac{\sum_i (\ln \tau_i - \ln \tau_i^{\rm pred})^2}{\sum_i (\ln \tau_i -  \langle \ln \tau_i \rangle)^2}
\end{equation}
where $\tau_i$ are the individual flow stress values within the set to be predicted, $\langle \ln \tau_i \rangle$ is their logarithmic average, and $\tau_i^{\rm pred}$ are the predicted values for $\tau_i$. Note that we use logarithmic quantities (i.e., we consider {\em relative} deviations) to ensure that all stress values in datasets which typically span 4-5 orders of magnitude are weighted equally (not doing so would imply that our performance measure is dominated by the largest stresses). 

For our entire set of data (simulation and experiment), a fit of equation (11) results in a high coefficient of determination ($R^2 = 0.958$). However, it is clear that the simple expression (11) is not the only possible form of a constitutive law that is consistent with scaling. For example, the transition regime between the asymptotic regimes might be represented by a modified scaling law which leaves the asymptotics unchanged, such as 
\begin{equation}
	T^{\beta}_{\rho} = \alpha^{\beta} + \dot{\Gamma}^{\beta} + \delta ( \dot{\Gamma}^{\beta})^{\eta}.
\end{equation}  
Fitting this law to all data produces a slightly improved fit ($R^2 = 0.967$) as shown in \figref{fig:fitting}. 

Importantly, Eq. (12) implies a different leading-order strain rate dependency of flow stress in the regime of low strain rates. Eq. (11), which has been used often in the literature, predicts that the strain rate increases linearly with the 'effective stress' $\tau_{\rm eff}^{\beta} = \tau^{\beta} - \alpha^{\beta} \mu b \sqrt{\rho^{\beta}}$, provided that the resolved shear stress exceeds the friction-like 'Taylor stress' : $\dot{\gamma} \propto \tau_{\rm eff}^{\beta}$ if $|\tau^{\beta}| \ge \alpha^{\beta} \mu b \sqrt{\rho^{\beta}}$. Eq. (12), on the other hand, predicts a nonlinear increase,  $\dot{\gamma} \propto (\tau_{\rm eff}^{\beta})^{1/\eta}$, as has been reported in simulations, see e.g. Miguel et. al. \cite{miguel2002dislocation}. 

\section{Machine learning}

In the following, we study the performance of different machine learning methods in predicting flow stresses based on a set of features which include dislocation density, strain rate, and strain as well as essential materials parameters (shear modulus, dislocation drag coefficient, Burgers vector length). The scaling analysis presented in the previous paragraph serves as a benchmark for prediction performance. All features and targets were logarithmically transformed and standardized before training, testing and prediction.

\subsection{ML methods}

To analyze the data, we use three different methods, namely kernel ridge regression (KRR), a decision tree, and a simple neural network. Kernel ridge regression is a memory based method that makes predictions for a new data point $x_{i}$ through its similarity to samples in the training set $x_{j}$. Similarity is quantified by a kernel function $k$ and a distance $d$ $k(d(x_{i},x_{j},\gamma)$, such that new predictions are made by a linear combinations of weighted kernel functions 
\begin{equation}
y_{i}=\sum_{j}^{N}w_{j}k(d(x_{i},x_{j},\gamma).
\end{equation}
$\gamma$ is here a generic kernel parameter. The weights $w$ are inferred by minimizing an $L_2$ regularized least squares problem which yields a closed form solution \cite{bishop}. We use the standard euclidean distance and the radial basis function kernel. The regularization parameter is varied between $10^{-5}$ and $10^{5}$ with one hundred logarithmically even spaced increments as well as the kernel parameter $gamma$. The combination with the best performance in the test set is chosen as final parameter set. Decision trees partition the feature space in greedy fashion and make predictions through the average value of training points within the partition to which a new data point $x_{i}$ is assigned. This partitioning is done in a sequential per-feature manner. The maximum tree  depth, minimum number of samples per partition and minimum number of samples to induce a split are tested with values $2^n, 1 \le n \le 5$ by exhaustive combination. The final model is a multilayer perceptron. Here the model is a set of stacked layers consisting of individual units/neurons with the nonlinear activation function $f(x)$. Each neuron $(i,j)$ is connected to all units $(i-1,k)$  of the previous layer via a weighted connection of weight $w$ and is further modified with a bias $b$. The intermediate value in the i-th layer on the j-th neuron then is given by 
\begin{equation}
z_{i,j}=f(\sum_{k}[b_{i,j} +  w_{k,j}z_{i-1,k}])
\end{equation} 
Weights and biases are trained in a stochastic gradient approach via backpropagation. In this work  the architecture is kept very small due to the limited number of samples: The structure is varied between two to four layer depth with a width of ten neurons. We test both the relu and sigmoid activation functions. Training is done with the Adam stochastic optimizer \cite{adam}. For more details on KRR, the interested reader is referred to the book of Bishop  \cite{bishop}, whereas for the other methods we refer to the standard textbook of Hastie \cite{hastie}. For KRR and the decision tree, the scikit-learn package was used \cite{scikit} while for the multilayer perceptron the Keras library was used with Tensorflow as backend \cite{keras,tensorflow}. 

\subsection{Training strategies and results}

\begin{table}[b]
	\caption{Values of coefficient of determination for the different prediction algorithms and training schemes.}
	\begin{tabular}{|l|l|l|l|l|l|l|l|l|}
		\hline
		learning & \multicolumn{2}{l|}{scaling fit}&\multicolumn{2}{l|}{kernel ridge} & \multicolumn{2}{l|}{decision tree} & \multicolumn{2}{l|}{perceptron} \\
		scheme & training & test & training & test & training & test & training & test \\\hline
		1 & 0.972 & 0.782 & 0.388 & - 0.45 &  0.879 & 0.430 & 0.989 & -2,32 \\
		2 & 0.972 & 0.782 & 0.961 & 0.514 & 0.963 & 0.237 & 0.988 & 0.202 \\
		3 & 0.967 & 0.944 & 0.988 & 0.958 & 0.990 &  0.927 & 0.958 &  0.946 \\\hline
	\end{tabular}
\end{table}

\begin{figure}[tb]
	\centering
	\hbox{}
	\includegraphics[width=0.9\textwidth]{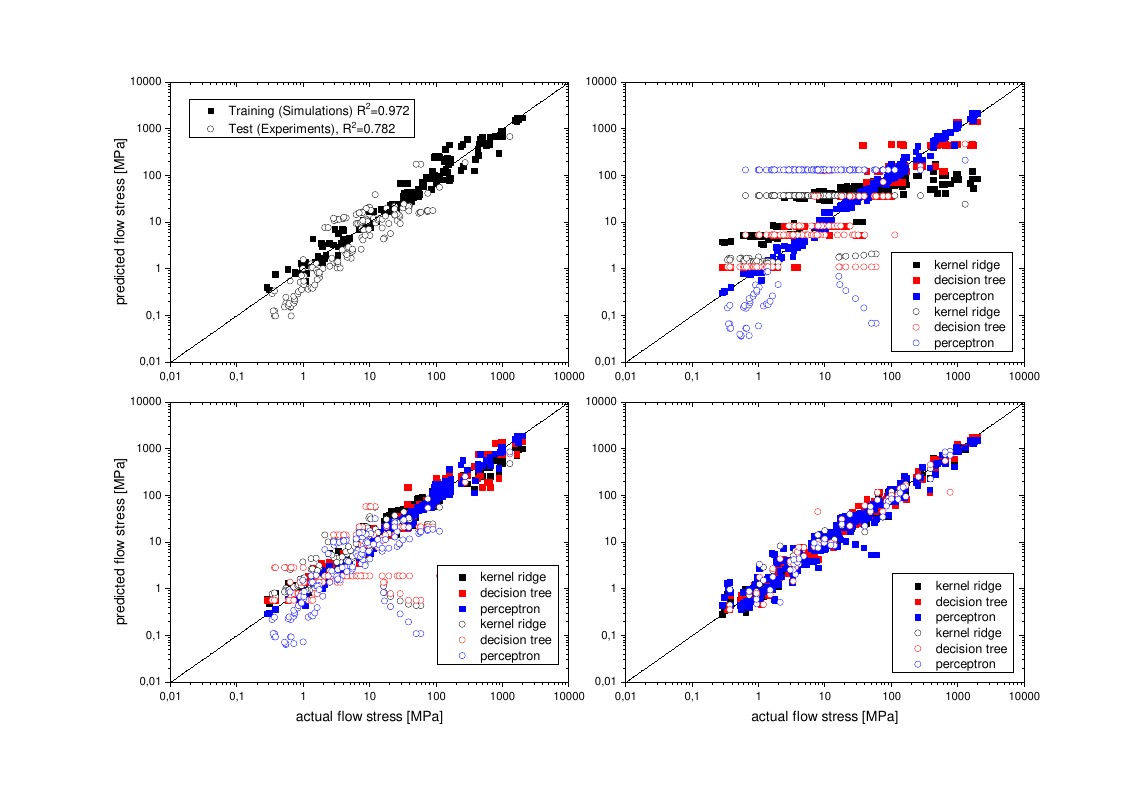}
	\caption{\label{fig:performance}
		Comparison of the performance of different prediction strategies.}
\end{figure}

We first consider a training strategy where we use the simulation data as training data and use the experimental data for testing (or validation). In this case,  a fit of the simulation data using Eq. (12) allows to reproduce the actual values with a coefficient of determination of $R^2 = 0.972$, i.e., the fit is slightly better than for the case considered in the previous section where {\em all} data were used for fitting. The same parameters then reproduce the experimental data with a value of $R^2 = 0.782$, indicating that the scaling analysis captures the experimental data well but also showing that the agreement is not perfect. A comparison of the scaling predictions and actual values for this training strategy is shown in \figref{fig:performance}, top left, using different symbols for predictions related to experimental and simulation data. 

The reduced reliability in predicting simulation data is to be expected, since the simulations represent highly idealized situations where deformation is controlled by dislocation interactions and dislocation drag alone, whereas the experiments are necessarily influenced by presence of other defects such as impurities or point defects, even though only experiments using single crystal specimen were considered. In line with this argument, deviations are strongest in the regime of low flow stresses, where the relative influence of such confounding factors is highest and accordingly the scaling prediction tends to underestimate the actual flow stress. 

If we apply the same training and validation strategy but replace the scaling fit by a machine learning algorithm, the results are at first glance quite disastrous (\figref{fig:performance}, top right). The $R^2$ compiled in Table 1 indicate that, while notably the perceptron algorithm is able to well reproduce the training (simulation) data, the performance of the trained algorithms in reproducing the test (experimental) data is either low ($R^2 < 0.5$) or non existing ($R^2 < 0$). 

In order to establish the reasons for this poor performance, we first look at the feature set provided to the algorithms. In this set, one variable (the plastic strain) is not actually used in the scaling analysis. This is in line with the materials physics idea that plastic strain is not a meaningful variable characterizing the internal state of a material. Of course, plastic strain can nevertheless be related to flow stress IF the initial state of the material (the initial dislocation density) is known, and evidently flow stress tends to increase with plastic strain in a strain hardening material. However, here the problem is exacerbated by the fact that the plastic strains in the dislocation dynamics simulations are quite small (typically 0.2\%) whereas in the experiments they may be much larger. Thus, the training (simulation) and test (experiment) data show poor overlap in feature space as far as this variable is concerned. 

If we remove plastic strain from the feature set, the predictive power of the ML algorithms increases (\figref{fig:performance}, bottom left). First of all, the 'training' performance is improved as most algorithms obtain better scores in reproducing the simulation data. Second, now all algorithms achieve a positive prediction score ($R^2 > 0$) for the test set, though their performance still falls below the performance of the scaling analysis. It is therefore fair to call the plastic strain in the present context a {\em confounding variable}.
	
One reason of poor prediction performance is the limited overlap of training and test data in feature space. Looking at \figref{fig:comparison}, we see that the parameter range covered by the simulations corresponds to an interval of lower $\Rho$ parameters than that of the experiments. The reason is that simulations are typically conducted at much higher strain rates than experiments. This is a simple consequence of the required effort: The computational effort to conduct a DDD simulation increases tremendously with decreasing strain rate, because the numerical stiffness of the simulations increases. The reason is that the simulation time step is controlled by the motion of fast nodes on close, strongly interacting dislocations, and therefore only weakly strain rate dependent, whereas the overall simulated time to reach a given strain is inversely proportional to the imposed strain rate. In experiment, the opposite is true: While a low strain rate of, say, $10^{-5}$ s$^{-1}$ is completely standard in a tensile test, achieving a high strain rate $> 10^4$ s$^{-1}$ in controlled test requires non standard equipment and significant effort.    

To resolve the problem of poor overlap of training and test data in feature space, we devise a third training strategy where training and test data are chosen randomly from the pool of {\em all} datasets, ensuring an approximately equal coverage of feature space by the training and test data. As seen in \figref{fig:performance}, bottom right, this leads to an improved performance which now matches the results of the scaling analysis. This is also manifest from the $R^2$ values compiled in Table 1 for this training scheme ('training scheme 3'): First, the improved overlap of training and test data ensures that also the scaling fit works better in reproducing the test set. Second, now all machine learning algorithms achieve comparable performance to the scaling analysis. 

\section{Discussion and Conclusions}

The present example is simple and surely not a critical test of the potential of machine learning approaches -- after all, we are dealing with the representation of a comparatively simple functional relationship in a low-dimensional parameter space. Nevertheless, it illustrates some of the pitfalls that may beset the interplay between high-troughput simulation, experiment, and machine learning. 

The first and obvious conclusion points to the necessity of ensuring adequate overlap between simulation and experiment, reflecting the observation that most machine learning approaches are better at intrapolation than at extrapolation \cite{webb2020learning}. This can be facilitated by carefully analysing the mathematical structure underlying the simulations: our scaling analysis actually demonstrates that a discrete dislocation dynamics simulation at high strain rate and high dislocation density may be equivalent to one at lower strain rate and lower dislocation density. Recognizing this fact can evidently accelerate the exploration of parameter space and allows to cover, at the same cost, a wider range of parameters. More generally speaking, it is helpful to exploit any symmetries in the mathematical formulation of the simulation problem, of which the scaling relations studied here are a nontrivial example. In our work on machine learning approaches to materials mechanics, we generally observe the importance of accounting for symmetries as well as symmetry breaking phenomena. Examles include the breaking of the translational symmetry of space due to strain localization in the run-up to creep failure \cite{biswas2020prediction}, and the use of feature functions of reduced symmetry for predicting bond breaking in statistically isotropic glasses under axial load \cite{font2022predicting}.

A further interesting point concerns the role of confounding variables, i.e., variables that (i) have no physical relation to the problem at hand, (ii) are systematically uneven distributed over the feature space such as to *suggest* a relationship with the other variables. It is ironic that, in the present study, the plastic strain plays the role of such a variable, given that most of mechanics of materials work is obsessed with fitting relationships between stress and (plastic) strain to empirical data, whereas few researchers ever bother to determine dislocation densities. 

If these caveats are addressed, our little study demonstrates that machine learning approaches can correctly infer relationships between simulation and/or experimental data and can thus be used to represent constitutive relationships governing material behavior. Because this representation is not governed by inherent biases such as physically unmotivated traditions regarding the structure of constitutive laws (see Eq. (\ref{eq:kocksmecking}), they may in fact do a better job than many human researchers \cite{hiemer2021mechanism}.

\section*{Declarations}
\subsection*{Competing interests}
  The authors declare that they have no competing interests.
\subsection*{Author's contributions}
S.H. carried out the machine learning analysis and wrote the section on machine learning, H.F provided the data base from high-throughput DDD simulations and experimental literature searches, performed the simulations and analyzed the simulation data. M.Z. and H.F. performed the scaling analysis, M.Z. drafted the manuscript. All authors corrected and edited the manuscript final version.
\subsection*{Funding}
M.Z. and S.H. acknowledge funding by the Deutsche Forschungsgemeinschaft (DFG) under Grants no. Za 171/13-1 and Za 171/15-1. S.H also acknowledges participation in the training activities of dhe DFG graduate school FRASCAL (GRK 2423/1). 

\subsection*{Acknowledgements}
Not applicable

\subsection*{Availability of data and material}
Not applicable

\bibliographystyle{bmc-mathphys} 
\bibliography{citescaling}   
%


\end{document}